\documentclass[%
reprint, 
superscriptaddress, 
showpacs,
amsmath, amssymb, 
aps, 
prl,
]{revtex4-1}
\usepackage{graphicx}
\usepackage{dcolumn}
\usepackage{bm}
\usepackage{hyperref}
\hypersetup{colorlinks=true, citecolor=blue, urlcolor=blue, linkcolor=blue}
\begin{document}
	\title{Cavity-enhanced and spatial-multimode spin-wave-photon quantum interface }
	\author{Minjie Wang}
	\author{Haole Jiao}
	\author{Jiajin Lu}
	\author{Wenxin Fan}
	\author{Zhifang Yang}
	\author{Mengqi Xi}
    \author{Shujing Li} 
    \email{lishujing@sxu.edu.cn}
	\author{Hai Wang}
	\email{wanghai@sxu.edu.cn}

	\affiliation{The State Key Laboratory of Quantum Optics and Quantum Optics Devices, Institute of Opto-Electronics, Shanxi University, Taiyuan 030006}%
	\affiliation{China Collaborative Innovation Center of Extreme Optics,
		Shanxi University, Taiyuan 030006, China}
	\date{\today}
	\begin{abstract}
Practical realizations of quantum repeaters require quantum memory simultaneously providing high retrieval efficiency, long lifetime and multimode storages. So far, the combination of high retrieval efficiency and spatially-multiplexed storages into a single memory remains challenging. Here, we set up a ring cavity that supports an array including 6 TEM$_{00}$ modes and then demonstrated cavity-enhanced and spatially-multiplexed spin-wave-photon quantum interface (QI). The cavity arrangement is according to Fermat’s optical theorem, which enables the six modes to experience the same optical length per round trip. Each mode includes “horizontal” and “vertical” polarizations. Via DLCZ process in a cold atomic ensemble, we create non-classically-correlated pairs of spin waves and Stokes photons in the 2*6=12 modes. The retrieved fields from the multiplexed SWs are enhanced by the cavity and the average intrinsic retrieval efficiency reaches $ \sim $ 70\% at zero delay. The storage time for the case that cross-correlation function of the multiplexed QI is beyond 2 reaches $ \sim $ 600$\mu$s.
\end{abstract}
\maketitle
\section{Introduction}
Entanglement distribution over a long distance via fibers is a critical task for quantum networks\cite{1,2,3,4}.The distance, over which one wants to directly distribute entanglement, is limited to the transmission losses in the fiber channel.Quantum repeaters (QRs) hold promise to overcome this loss problem\cite{5, 6}.QRs\cite{7} divide the long distances into a number of elementary links, each of which is composed of two nodes.Entanglement is first generated in each link and then stored the two nodes.Various physical systems\cite{8}, such as atomic ensembles \cite{5, 6, 9} and single quantum systems, including single atoms\cite{10, 11}, ions\cite{12, 13} and solid-state spins\cite{14, 15} have been demonstrated as nodes. 

The well-known Duan-Lukin-Cirac-Zoller (DLCZ) protocol use atomic-ensemble-based quantum memories (QMs) as repeater nodes\cite{5, 6}.For establishing entanglement in a link, write pulses are applied into two atomic ensembles that are belong to a link, which creates non-classically-correlated pairs or entangled of a spin wave (SW) and a Stokes photon via spontaneous Raman emissions in individual ensembles.The SWs are stored in the ensembles (nodes) and the photons are sent to the central station between the nodes for Bell measurements.When entanglement is established in each link, the SWs are converted into photons for entanglement swapping between adjacent links\cite{5, 6}.Significant progress has been achieved on experimental generations of non-classically correlated or entangled pairs of a photon and a SW via DLCZ protocol\cite{16,17,18,19,20,21,22,23,24,25,26,27,28,29,30,31,32,33,34,35,36}.Alternately, the non-classically correlated or entangled pairs can be achieved by using SPDC process together with quantum memories\cite{37,38,39,40,41}, where, the quantum memories are realized via “absorptive” (“read-write”\cite{42}) storage schemes that are based on electromagnetically induced transparency(EIT)\cite{37}, atomic frequency comb (AFC)\cite{38,39,40}, and off-resonant cascaded absorption\cite{41} \textit{et.al.}\ Realization of QRs in practice requires memories to have high performance characteristics, including high retrieval efficiency, long lifetimes, and multi-mode storage capacity\cite{6,8}.

To realize long-lived QMs, significant progresses have been made with cold atomic ensembles\cite{20, 21, 23, 24, 33, 35, 43}.These studies show that atomic-motion-induced decoherence can be suppressed either by lengthening spin-wave wavelengths\cite{20, 35, 43} or confining the atoms in optical lattices\cite{21, 23, 24, 33}.Inhomogeneous-broadening-induced decoherence may be suppressed by storing SWs in magnetic-field-insensitive (MFI) coherences\cite{20, 21, 23, 24, 33, 35}.High-efficiency memories for optical quantum states have been achieved by using either high-optical-depth atomic ensembles\cite{43,44,45,46,47,48,49} or coupling moderate-optical-depth atoms to a low finesse optical cavity\cite{32, 33, 35, 50,51,52,53}.With high optical-depth cold atoms, the efficiencies of EIT storages of single photon\cite{46} and polarization qubit\cite{45} reach to 85\% and 68\%, respectively, together with a lifetimes of $ \sim $ 10$\mu$s. Using the cavity-enhanced scheme, high-retrieval-efficiency and long-lived atom-photon QI have been demonstrated with optical-lattice cold atoms\cite{24}via DLCZ protocol, where zero-delay retrieval efficiencies reaches 76\% and 1/e lifetimes 220 ms.However, those experiment only can store single modes.It has been pointed out that QRs using single-mode memory have very slow repeater rates for practical use\cite{54,55,56,57}.To overcome this problem, multiplexed QR schemes were proposed\cite{36,55,56,57,58}. When QR use multiplexed QI storing \textit{N}\ modes as nodes, the entanglement creation rate per link will be increased by the factor of \textit{N}\cite{57}. 

AFC quantum memories stored REID crystals have large bandwidth and are suitable for temporally and spectrally-multiplexed storages\cite{4}.In the past decades, significant progresses along the directions have been experimentally made\cite{28, 29, 40, 58}.Recently, non-classical correlations atom-telecom-photon QI with 1250 temporally-multiplexed modes\cite{59} has been demonstrated with REID crystals.Also, a multiplexed quantum teleportation from a telecom photon to a solid-state qubit is achieved\cite{60}. 

DLCZ quantum memories in gas-state atomic ensembles are suitable for spatially-multiplexed storages.By collecting Stokes fields in multiple spatial modes, multiplexed  QI that generates spin-wave-photon entanglement in six modes have been demonstrated with a cold atomic ensemble\cite{36}.On the basis of that work\cite{36}, a long-lived and spatially-multiplexed atom-photon entanglement QI has been demonstrated\cite{61}.Pu \textit{et.al.}\ realized a multiplexed DLCZ-type quantum memory with 225 individually accessible memory cells using a cold atomic ensemble\cite{34}.Massively-multiplexed DLCZ-type quantum memories have been demonstrated in cold atoms via spatially-resolved single-photon detection\cite{62}.Temporally-multiplexed DLCZ QIs have been also demonstrated by applying a train of write pulses each of which along different directions to a cold atomic ensemble\cite{63} or applying a reversible gradient magnetic field to a cold atomic ensemble to control the rephasing of the spin waves\cite{64}.Since the temporally-multiplexed QIs emit write-out and read-out photons in a specific spatial mode, the coupling of the photons to a ring cavity is easy to achieve.Recently, the cavity-enhanced retrieval\cite{65}, cavity-enhanced noise suppression\cite{66}, and spin-wave multiplexed atom-cavity electrodynamics\cite{67} have been experimentally demonstrated with the temporally-multiplexed DLCZ quantum memories.

Spatial multiplexing is particularly powerful in QRs because it also reduces the demands on memory lifetime\cite{6,56}.It has been used to demonstrate quantum communication between two nodes\cite{68}, or hardware quantum repeater nodes\cite{69}.However, in those experiments, intrinsic retrieval efficiencies are lower ($\le10\% $ ).It is important to pave the way to enhance retrieval efficiency of the spatially-multiplexed QIs via cavity-enhanced scheme since the scheme is suitable for low-optical-depth atoms, for example, optical-lattice atoms, which hold promise for achieving sub-second lifetime storages.However, the realization of cavity-enhanced scheme in spatially-multiplexed QI is difficult since such QIs emit write-out and read-out photons in spatial modes with different directions.

Here, we overcome this difficulty by establishing a ring cavity that supports an array of TEM$_{00}$ modes, which promise us to demonstrate high-efficiency and spatially-multiplexed spin-wave-photon quantum interface (QI) in a cold atomic ensemble.The ring cavity is formed by four mirrors and inserted with two pairs of optical lenses, each of which is setup in confocal manner.A cold atomic ensemble is placed at the center of one of pairs of the lens, which promises us to achieve an effective atom-photon interaction.The cavity arrangement enables the mode array to be recovered after a round trip, meaning that the cavity supports the mode array.The presented mode array includes six optical TEM$_{00}$ modes, each of the modes includes “horizontal” and “vertical” polarizations. Via DLCZ process, we create non-classically-correlated pairs of spin waves and Stokes photons in the \textit{N}\ = 2*6 =12 modes.The cavity set is also based on Fermat's optical theorem, thus, the modes of the array experience the same optical length.So, the retrieved fields from the multiplexed SWs are simultaneously resonant with the cavity, which enable cavity-enhanced retrievals. 

\section{Experimental method}
The experimental setup is shown in Fig.1(a), where the ring cavity is a vertical view. The cavity is formed by three high-reflection mirrors HR$_{1,2,3}$ and an output coupler (OC) with partial reflection. To achieve cavity-enhanced and spatially-multiplexed atom-photon interface, we set up the cavity that can support multiple TEM$_{00}$ spatial modes. At first, we couple six TEM$_{00}$ modes into the cavity through OC. The modes propagate parallel just after OC, which are arranged in an array as shown in Fig.1d. The modes are named \textit{C}$1$, \textit{C}$2$ …\textit{C}$6$, respectively, each has a very small divergence angle of ${\rm{0}}{\rm{.0}}{{\rm{5}}^{\rm{o}}}$  and \textit{H}- and \textit{V}- polarizations. We describe the cross section of the array at Fig.1d as ${A_{{\rm{1d}}}} = \left[ {\begin{array}{*{20}{c}}
		{C1}\\
		{{C3}}\\
		{{C5}}
	\end{array}\begin{array}{*{20}{c}}
		{C2}\\
		{{C4}}\\
		{{C6}}
\end{array}} \right]$. In the vertical view of the cavity, the modes \textit{C}$1$, \textit{C}$3$ and \textit{C}$5$ (\textit{C}$2$, \textit{C}$4$ and \textit{C}$6$) overlap. So, we only plot two lines to represent [\textit{C}$1$, \textit{C}$3$ and  \textit{C}$5$], [\textit{C}$2$, \textit{C}$4$ and \textit{C}$6$] modes. We now consider the case that the modes start from Fig.1d and propagate along clockwise.After the mirror HR$_1$, the array experiences a mode-swapping operation in plane and is transformed into ${A_{{\rm{1e}}}} = \left[ {\begin{array}{*{20}{c}}
{{C2}}\\
{{C4}}\\
{{C6}}
\end{array}\begin{array}{*{20}{c}}
{{C1}}\\
{{C3}}\\
{{C5}}
\end{array}} \right]$, see Fig.1e. For extending the number of spatial modes interacting with the atoms, we inserted a polarization interferometer in the cavity. The interferometer is formed by two beam displacers BD1 and BD2, which has been used in our previous work\cite{35}. As shown in Fig.1a, the \textit{H}- and \textit{V}- polarizations of each optical mode, for example, \textit{C}$i$, will be split into two arms, \textit{C}$i_{H}$ and \textit{C}$i_{V}$. So, after BD1, the number of the modes is extended to 2*6 =12, seeing Fig.1f for details. To achieve effective atom-photon interactions, we focus the mode array at the center of the atoms (z=0). For which, we insert a pair of lens L1 and L2 that are arranged in a confocal manner in the interferometer. The 6 pairs of \textit{H}-polarization (\textit{C}$i_{H}$) and \textit{V}-polarization arms (\textit{C}$i_{V}$) go through L1 and are focused a light spot at z=0. After L2, the modes parallel propagate and then go through BD2. BD2 combine two arms of each mode, for example, \textit{C}$i_{H}$ and \textit{C}$i_{V}$ arms into single spatial mode \textit{C}$i$, which include two polarization modes. After the BD2, the mode array, which represents the cross-section at Fig.1g, is transformed into ${A_{{\rm{1g}}}} = \left[ {\begin{array}{*{20}{c}}
{{C5}}\\
{{C3}}\\
{{C1}}
\end{array}\begin{array}{*{20}{c}}
{{C6}}\\
{{C4}}\\
{{C2}}
\end{array}} \right]$. From ${A_{{\rm{1g}}}}$, one can see that the confocal device makes a dialogue-swapping operation on the array, which is different from the operation of the mirrors. The swapping of the mode positions in the array will result in cross-talk between the modes and then degrade quantum correlation of the spin-wave-photon pairs. For eliminating the dialogue swapping, we insert another confocal device formed by lenses L3 and L4. As shown in Fig.1a, the multiplexed modes are focused at HR$_3$ after the lens L3. After the confocal device together with the mirror HR$_3$, the mode array is transformed into ${A_{{\rm{1h}}}} = \left[ {\begin{array}{*{20}{c}}
{{C2}}\\
{{C4}}\\
{{C6}}
\end{array}\begin{array}{*{20}{c}}
{{C1}}\\
{{C3}}\\
{{C5}}
\end{array}} \right]$ at the cross-section Fig.1h. After OC, the mode positions of the array will be transformed back the original one as shown in Fig.1e. Thus, the multiplexed modes are self-reproduction and are supported by the cavity. For realizing cavity-enhanced retrievals of the multiplexed QI, the multiplexed modes \textit{C}$i_{\alpha\ }$ (i=1 to m, $\alpha $=\textit{H}\ , \textit{V}\ ) are required to be simultaneously resonant with the cavity. Our presented cavity arrangement satisfies this requirement. As shown in Fig.1a, at point z=0, the multiplexed fields emit from a light spot. According to Fermat's optical theorem\cite{70}, when each of the modes \textit{C}$i_{H}$ or \textit{C}$i_{V}$ (i=1 to m) propagate from z=0 to z=0 point per round trip, it will experience a minimum optical length. Thus, the optical lengths of the modes \textit{C}$i_{H}$ or \textit{C}$i_{V}$ are the same. The relative phase between the two arms of the interferometer, i.e., the cavity mode \textit{C}$i_{H}$ and \textit{C}$i_{V}$ , can kept to be $2m\pi $  by stabilizing the temperatures of the two BDs, where m is integer. In the presented experiment, we couple a locking laser beam to \textit{C}$1_{H}$ mode through OC. Its leaks from HR$_3$ are used to stabilize the cavity length. With the stabilization, we may make the multiplexed modes be simultaneously resonant with the cavity. For each cavity mode, the transmission loss and transmission at OC are the same. So, the finesses of the 12 cavity modes are all $ \sim $ 16, which is the same that in Ref.\cite{35}.
\begin{figure*}[ht]
	\centering
	\includegraphics[width=5.5in]{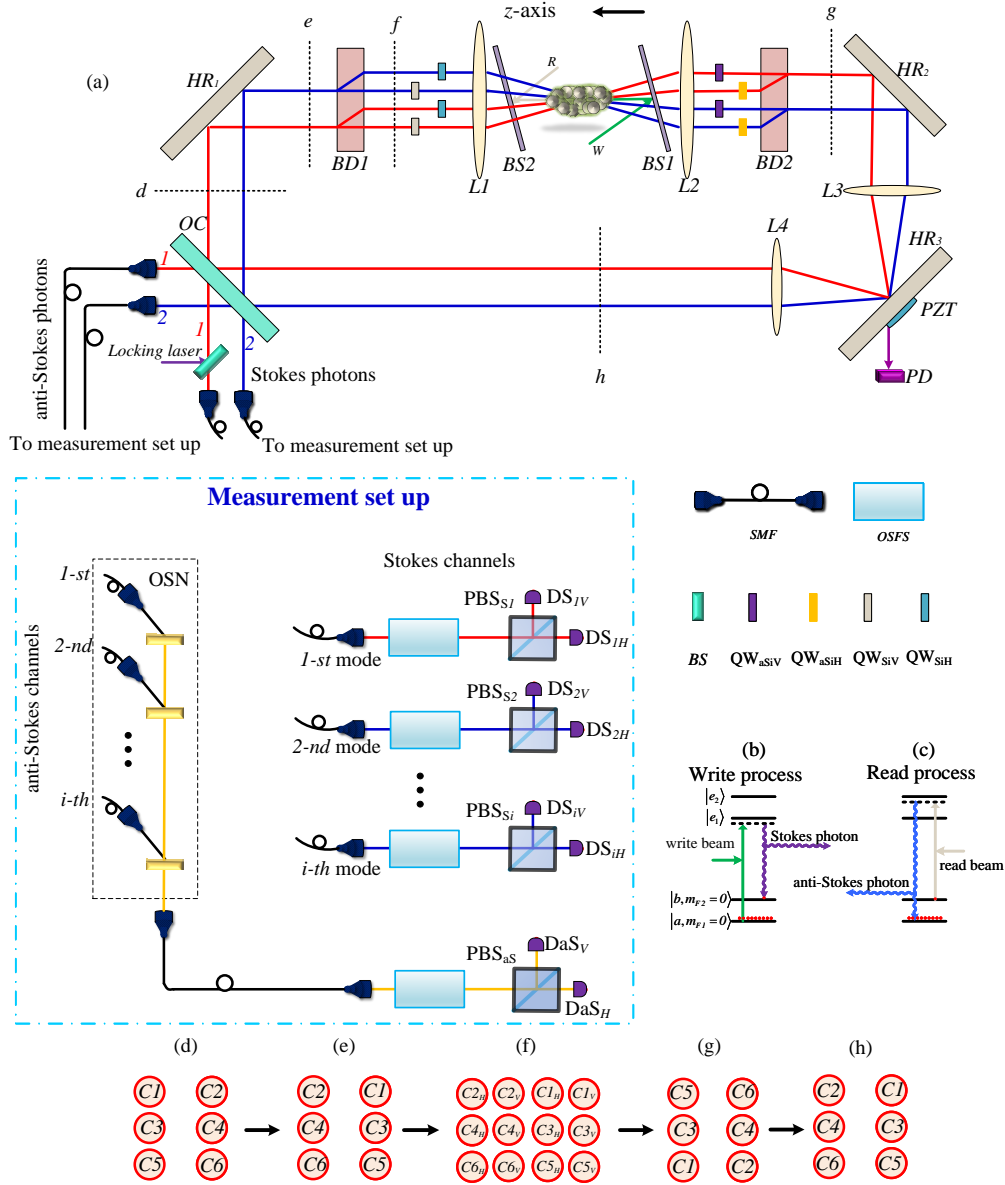}
	\caption{Schematic diagram for experimental setup. (a): The ring cavity is formed by three flat mirrors (HR$_{1,2,3}$) with high reflection and an output coupler (OC) with a reflectance of 80\%. A polarization interferometer consists of two beam displacers (BD1 and BD2) is inserted into a ring cavity. Two pairs of lenses (L1 and L2, L3 and L4) are inserted in the cavity for spatially constraining the modes. Each pair of the lenses forms a confocal construction, and the focus is on the atomic center and HR$_3$ with piezoelectric transducer (PZT). The locking laser pulse is coupled to the cavity through a beam splitter (BS). Leaks of the cavity-locking pulse from HR$_3$ are detected by a fast photodiode (PD) to generate error signals. The error signals are amplified and used to drive a piezoelectric transducer (PZT) to stabilize the cavity length. OSN: optical switch network; OSFS: optical-spectrum-filter set; BS1 (BS2): non-polarizing beam splitter with a reflectance of 1\% (3\%). (b) and (c) are relevant atomic levels. W (R): write (read) laser. (d): the mode section distribution after OC. (e): the mode section distribution after HR$_1$. (f): the mode section distribution after BD1. (g): the mode section distribution after BD2. (e): the mode section distribution after L4. }.
	
	\label{figure1}
\end{figure*}

The atomic ground levels  $\left| a \right\rangle  = \left| {5{S_{1/2}},F{\rm{ = 1}}} \right\rangle $ and  $\left| b \right\rangle  = \left| {5{S_{1/2}},F{\rm{ = 2}}} \right\rangle $ together with the excited level $\left| {{e_{\rm{1}}}} \right\rangle  = \left| {5{P_{1/2}},F'{\rm{ = 1}}} \right\rangle $  ($\left| {{e_{\rm{2}}}} \right\rangle  = \left| {5{P_{1/2}},F'{\rm{ = 2}}} \right\rangle $) form a  $\Lambda $-type system [Fig.1(b) and(c)]. We apply a bias magnetic field (4G) along z-direction to define quantum axis. After the atoms are prepared in the Zeeman state $\left| {a,{m_{{F_a}}} = 0} \right\rangle $,we start the generations of the spin-wave-photon pairs. Similar to that in our previous work\cite{35}, we apply a 795-nm right-circularly-polarized write pulse with red-detuned by 110 MHz to the $\left| a \right\rangle  \to \left| {{e_{\rm{1}}}} \right\rangle $  transition to the atoms along z-axis through a beam splitter BS1 at the beginning of a trial. This write pulse drives ${\sigma ^ + }$-transition $\left| {a,{m_{{F_a}}} = 0} \right\rangle  \leftrightarrow \left| {{e_1},{m_{Fe}}_{_1} =  + {\rm{1}}} \right\rangle $ and induces Raman transition $\left| {a,{m_{Fa}} = {\rm{0}}} \right\rangle  \to \left| {b,{m_{Fb}} = {\rm{0}}} \right\rangle $ [Fig. 1(b)], which emit Stokes photons on ${\sigma ^ + }$-polarized $\left| {b,{m_{{F_b}}} = 0} \right\rangle  \leftrightarrow \left| {{e_1},{m_{F{e_1}}} =  + {\rm{1}}} \right\rangle $ and simultaneously create SW excitations stored in the clock coherence $\left| {a,{m_{Fa}} = {\rm{0}}} \right\rangle  \to \left| {b,{m_{Fb}} = {\rm{0}}} \right\rangle $.The Stokes photons propagate along z-axis and are right-circularly polarized.If the Stokes photon is emitted into the cavity mode \textit{C}$1_{H}$, \textit{C}$1_{V}$, \textit{C}$2_{H}$, \textit{C}$2_{V}$, …, \textit{C}$6_{H}$, or \textit{C}$6_{V}$, one collective excitation will be created in the SW mode \textit{M}$1_{H}$, \textit{M}$1_{V}$, \textit{M}$2_{H}$, \textit{M}$2_{V}$, …, \textit{M}$6_{H}$, \textit{M}$6_{V}$ defined by the wave-vector $k_{MiH(V)}^{} = {k_W} - k_{SiH(V)}^{}$, where ${k_W}$ is the wave-vector of the write pulse and $k_{S{_{iH(V)}}}^{}$ that of the Stokes photon propagating in the cavity $C{i_{H(V)}}$  mode. The angle between any two adjacent modes (for example, the mode pairs [\textit{C}$1_{H}$, \textit{C}$1_{V}$] and [\textit{C}$2_{H}$, \textit{C}$2_{V}$], and et. al.) is ${\theta _R} \approx {\rm{0}}{\rm{.2}}{{\rm{1}}^o}$, which is beyond the angular-separation lower bound required for independent storages of adjacent spin-wave modes\cite{61}. As shown in Fig.1(b), the write pulse also induces the Raman transition $\left| {a,{m_{Fa}} = {\rm{0}}} \right\rangle  \to \left| {b,{m_{Fb}} = {\rm{2}}} \right\rangle $ via $\left| {{e_1},{m_{Fe}} = {\rm{1}}} \right\rangle $, which emit left-circularly-polarized Stokes photons into any cavity modes and simultaneously create magnetic-field-sensitive ($\left| {{m_{Fa}} = {\rm{0}}} \right\rangle  \leftrightarrow \left| {{m_{Fb}} = {\rm{2}}} \right\rangle $) SWs. However, the left-circularly-polarized Stokes photons moving toward right are excluded by BD1\cite{35}. In our presented experiment, the excitation probabilities for the various spatial modes are small and basically symmetrical, i.e., ${\chi _{{1_H}}} = {\chi _{{1_V}}} = ... = {\chi _{{i_\alpha }}} = ...{\chi _{{6_V}}} <  < {\rm{1}}$($\alpha  = H,V$), the joint state of the spin-wave-photon pair in ${i_\alpha }$-th mode can be written

${\left| \psi  \right\rangle _{i\alpha }} = {\left| {\rm{0}} \right\rangle _{Si\alpha }}{\left| {\rm{0}} \right\rangle _{Mi{\alpha _{}}}} + \sqrt {{\chi _{i{\alpha _{}}}}} {\left| {\rm{1}} \right\rangle _{Si{\alpha _{}}}}{\left| {\rm{1}} \right\rangle _{Mi\alpha }} + O({\chi _{i{\alpha _{}}}})$
where, ${\left| {\rm{1}} \right\rangle _{S{_i\alpha }}}$(${\left| {\rm{1}} \right\rangle _{M{i_\alpha }}}$) denotes a photon (a collective excitation) in ${i_\alpha }$-th Stokes (spin-wave) mode, ${\left| {\rm{0}} \right\rangle _{S{i_\alpha }}}$(${\left| {\rm{0}} \right\rangle _{M{i_\alpha }}}$) the vacuum state.The Stokes fields in \textit{C}$i_{H}$ and \textit{C}$i_{V}$ modes are combined after BD1 and direct into the cavity mode \textit{C}${i}$.

After a storage time \textit{t}\, we apply a right-circularly-polarized read pulse with a red-detuned 110MHz to the transition $\left| b \right\rangle  \leftrightarrow \left| {{e_2}} \right\rangle $ to the atoms.The read pulse counter-propagates with the write beam, which drive the ${\sigma ^ - }$-transition $\left| {b,{m_{{F_b}}} = 0} \right\rangle  \leftrightarrow \left| {{e_2},{m_{F{e_2}}} =  - {\rm{1}}} \right\rangle $ and convert the SW ${\left| {\rm{1}} \right\rangle _{Mi\alpha }}$ into an anti-Stokes photon. The anti-Stokes photon is collectively emitted on ${\sigma ^ - }$-transition $\left| {a,{m_{{F_a}}} = 0} \right\rangle  \leftrightarrow \left| {{e_2},{m_{F{e_2}}} =  - {\rm{1}}} \right\rangle $, whose wave vector is determined by the phase-matched condition $k_{aSi{\alpha _{}}}^{} = {k_w} - k_{Si\alpha }^{} + {k_r} \approx  - k_{Si\alpha }^{}$, where, ${k_w}$ (${k_r}$) and $k_{Si\alpha }^{}$ ($k_{aSi{\alpha _{}}}^{}$) denote the wave vectors of the write (read) laser pulse and Stokes (anti-Stokes) photon, respectively. The anti-Stokes photon retrieved from the SW ${Mi_{\alpha }}$ propagates along the direction opposite to the Stokes photons in ${Ci_{\alpha }}$ modes and is also right-circularly polarized. The anti-Stokes photon ${Ci_{H}}$ (${Ci_{V}}$) is transformed into the \textit{H} (\textit{V})-polarized photon by a $\lambda$/4 wave plate named QW$_{aSiH}$ (QW$_{aSiV}$).After BD2, they are combined into \textit{C}${i}$ mode and are supported by the cavity. The Stokes (anti-Stokes) photons in the various modes are produced (retrieved) by a write (read) field, the frequencies of Stokes or anti-Stokes are the same. In our presented, the length of the mode \textit{C}$1_{H}$ is actively stabilized by using a locking beam. By tuning the frequency of the write (read) laser beam, the Stokes (anti-Stokes) photons in the various cavity modes are simultaneously resonate with the cavity. So, the retrieval efficiencies of the various modes are enhanced through Purcell effect\cite{71}. As shown in Fig. 1(a), the escaping Stokes photon from OC in \textit{C}${i}$ mode is coupled to a single-mode fiber SMF$_{Si}$ and then is guided into polarization-beam splitter PBS$_{Si}$. The two outputs of PBS$_{Si}$ are sent to the single-photon detectors DS$_{iH}$ and DS$_{iV}$. The escaping anti-Stokes photons from OC in \textit{C}${i}$ mode is coupled to a single-mode fiber SMF$_{aSi}$. To enable our presented multiplexed QI to be available for the multiplexed QR scheme, we route the retrieved (anti-Stokes) fields in the SMF$_{aSi}$ (i= 1 to 6), each include \textit{H}- and \textit{V}- polarization, into a common single-mode fiber by using an optical switch network (OSN)\cite{36}. Passing through the common single-mode fiber (CSMF), the retrieved (anti-Stokes) fields direct into a polarization splitter labeled as PBS$_{aS}$. The two outputs of PBS$_{aS}$ are sent to single-photon detectors DaS$_H$ and DaS$_V$. Before each polarization splitter, we placed optical-spectrum-filter sets to suppress noise from the write/read beams, and leakage from locking beam.

\section{Experimental results}
The presented multiplexed QI can store $N = 2 \times m$  SW modes and then increase the probability of generating a spin-wave-photon pair per trial by the factor of $N$ , where \textit{m} (up to 6) is the spatial mode number. The probability increases by multiplexed QIs have been demonstrated in our previous works \cite{36, 61}. In the presented work, we measured the probability of generating a spin-wave-photon pair per trial as the function of the stored mode number and presented the measured results in the supplementary material. 

To show that cavity enhances the retrievals of the multiplexed QI, we measured the retrieval efficiencies of the multiplexed SWs.The intrinsic retrieval efficiency of the spatially-multiplexed QI storing$N=2\times m=12$ SW modes is defined by$R_{inc}^{(N = 12)} = \sum\limits_i^{m = 6} {\frac{{\left( {P_{S,aS}^{{(i_{H}})} + P_{S,aS}^{{(i_{V}})}} \right)}}{{{\eta _{aS}}\left( {{P_{D{S_{iH}}}} + {P_{D{S_{iH}}}}} \right)}}} $,where, $P_{S,aS}^{{(i_{H}})}$ ($P_{S,aS}^{{(i_{V}})}$) denotes the probability of detecting a coincidence between the detectors DS$_{iH}$ (DS$_{iV}$) and DaS$_{H}$ (DaS$_{V}$) per trial,${\eta _{{\rm{a}}S}} = {\eta _{cav}}{\eta _t}{\eta _D} \approx 14\% $ is the total detection efficiency of read-out (anti-Stokes) channel, which includes the efficiency of light escaping from the ring cavity${\eta _{esp}} \approx {\rm{60\% }}$, the cavity-to-detector transmission efficiency${\eta _t} \approx {\rm{34\% }}$, the efficiency of the single-photon detectors${\eta _D} \approx {\rm{68\% }}$.
It is found that in our presented experiment, the intrinsic retrieval efficiency can be simply written as $R_{inc}^{(N = 12)} \approx \frac{{\sum\limits_{i = {\rm{1}}}^6 {\left( {R_{inc}^{{(i_{H}})} + R_{inc}^{{(i_{V}})}} \right)} }}{N} = \bar R_{inc}^{(N = {\rm{12)}}}$, where, $R_{inc}^{{(i_{\alpha }})} = \frac{{P_{D{S_{i\alpha }},a{S_\alpha }}^{}}}{{{\eta _{aS}}P_{D{S_{i\alpha }}}^{}}}$  denotes the intrinsic retrieval efficiency of the ${i_\alpha }$ -th SW mode,$\bar R_{inc}^{({\rm{12}})}$  the average intrinsic retrieval efficiency of the multiplexed SW modes.The above equation shows that the retrieval efficiency of the multiplexed QI, The square dots in Fig.2 plot the measured efficiency  $R_{inc}^{(N = 12)}$ as a function of storage time t. The solid curves are the fittings to the retrieval efficiencies according to the function of$R_{inc}^{(N = 12)}(t) = R({\rm{0}})\left( {\exp \left( {{{{\rm{ - }}{{\rm{t}}^2}} \mathord{\left/
				{\vphantom {{{\rm{ - }}{{\rm{t}}^2}} {\tau _0^2}}} \right.
				\kern-\nulldelimiterspace} {\tau _0^2}}} \right){\rm{ + }}\exp \left( {{{{\rm{ - t}}} \mathord{\left/
				{\vphantom {{{\rm{ - t}}} {{\tau _0}}}} \right.
				\kern-\nulldelimiterspace} {{\tau _0}}}} \right)} \right)/2$ , which yield zero-delay retrieval efficiencies  $R({\rm{0}}) \approx 70\% $ and the lifetimes ${\tau _0} = 600\mu s$ , respectively.

\begin{figure}[h]
	\centering
	\includegraphics[width=3in]{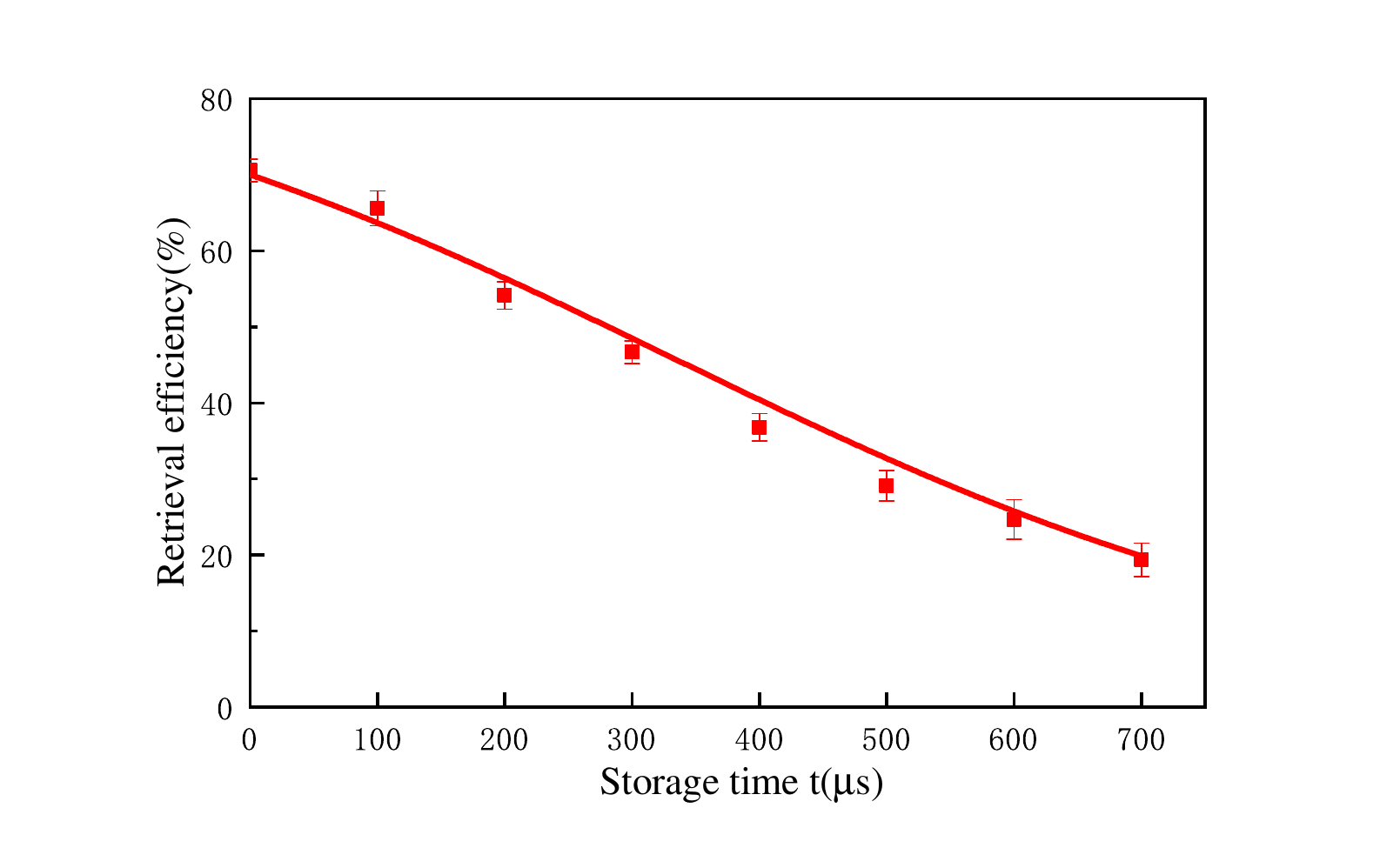}
	\caption{The measured retrieval efficiency $R_{inc}^{(N = 12)}$ (red squares) of the spatially-multiplexed QI as a function of the storage time t. The red curve is the fitting to the measured data according to $R_{inc}^{(N = 12)}(t) = R({\rm{0}})\left( {\exp \left( {{{{\rm{ - }}{{\rm{t}}^2}} \mathord{\left/
						{\vphantom {{{\rm{ - }}{{\rm{t}}^2}} {\tau _0^2}}} \right.
						\kern-\nulldelimiterspace} {\tau _0^2}}} \right){\rm{ + }}\exp \left( {{{{\rm{ - t}}} \mathord{\left/
						{\vphantom {{{\rm{ - t}}} {{\tau _0}}}} \right.
						\kern-\nulldelimiterspace} {{\tau _0}}}} \right)} \right)/2$ , yielding R(0)=70\%, ${\tau _0} = 600\mu s$.}.
	
	\label{figure4}
\end{figure}
Next, we measure the cross-correlation function between the Stokes and anti-Stokes of the multiplexed QI with N=12 mode storages. It is defined by $g_{S,aS}^{(N = {\rm{12}})} = \frac{{\sum\limits_{i = 1}^6 {\left( {P_{S,aS}^{{(i_{H}})} + P_{S,aS}^{{(i_{V}})}} \right)} }}{{\sum\limits_{i = 1}^6 {\left( {{P_{D{S_{iH}}}}P_{aS}^{{(i_{H}})} + {P_{D{S_{iH}}}}P_{aS}^{{(i_{V}})}} \right)} }}$ , which is simply expressed as a [Supplementary material]:
\begin{equation}
	g_{S,aS}^{(N = {\rm{12}})} \approx {\rm{1}} + \frac{{\chi R{{_{inc}^{(12}}^)}(t)}}{{\chi R{{_{inc}^{(12}}^)}(t) + \chi \left( {1 - R{{_{inc}^{(12}}^)}(t)} \right){\xi _{se}}A + Z}}
\end{equation}

where, \textit{Z} denotes background noise for any of the anti-Stokes detection channels, with ${Z^{{(1_{H}})}} = {Z^{{(1_{V}})}} = {Z^{{(2_{H}})}}... = {Z^{{(6_{V}})}}$ assumed, A = 2F/$\pi$  is the cavity-enhanced factor, F is the cavity finesses for any cavity modes, with  ${F^{({{\rm{1}}_H})}} = {F^{({{\rm{1}}_V})}} = {F^{({{\rm{2}}_H})}}... = {F^{({{\rm{6}}_V})}} \approx {\rm{16}}$ in the presented experiment, ${\xi _{se}}$  denotes the branching ratio corresponding to the read-photon transitions\cite{72}.The red square dots in Fig.3 plots the measured $g_{S,aS}^{(N = 12)}$  as a function of the storage time t. The red curve is the fitting to the measured results of $g_{S,aS}^{(N = 12)}$  according to Eq.(1), where, the data of $\bar R_{inc}^{({\rm{12}})}(t)$  are taken from the Fig.2. Fig.3 shows that the cross-correlation function $g_{S,aS}^{(N = 12)}$  decays with t, which result from that the retrieval efficiencies decay with t. From Fig.(3), one can see that  $g_{S,aS}^{(N = 12)}$ is well above 2 at $t = 600\mu s$ . 
\begin{figure}[h]
	\centering
	\includegraphics[width=3in]{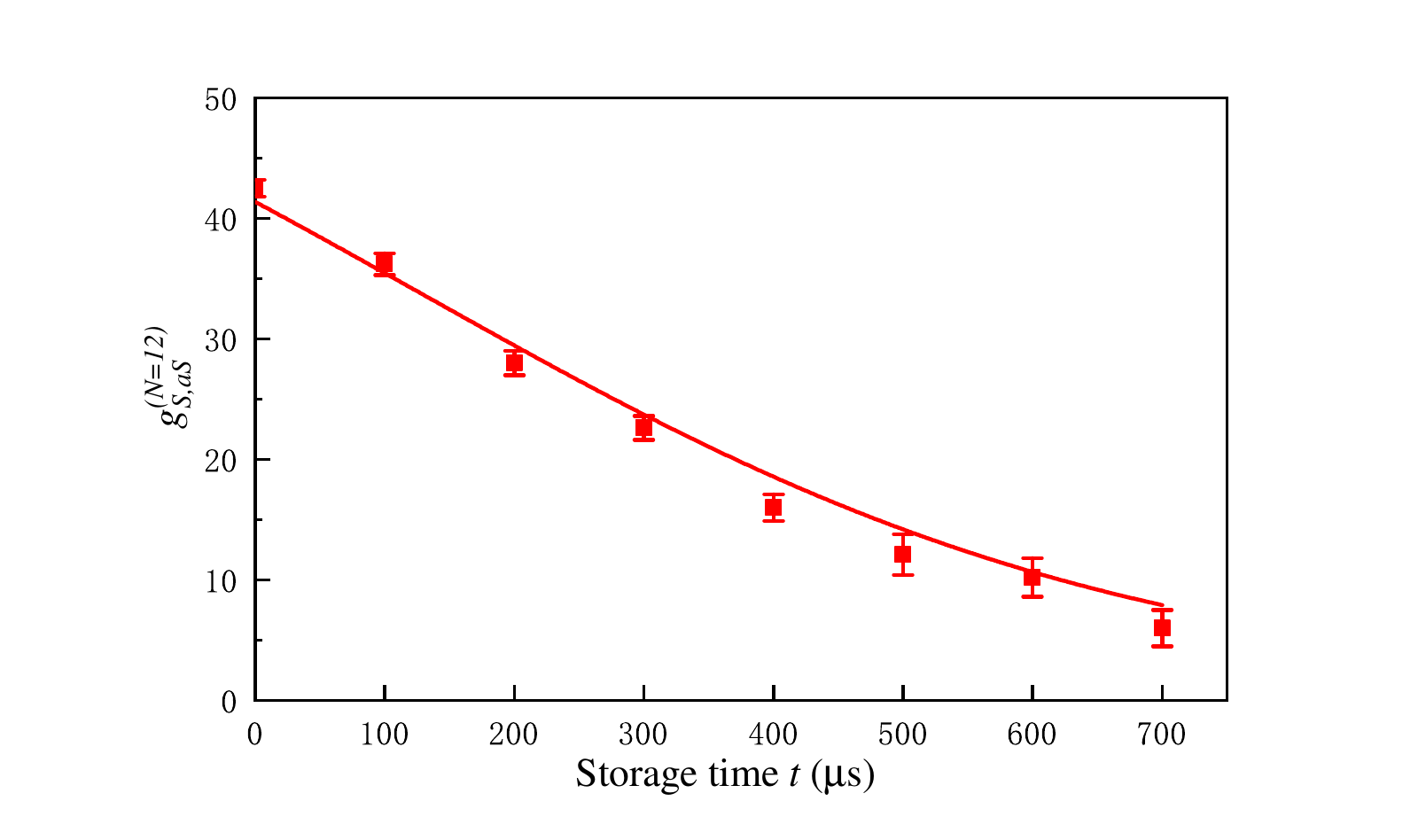}
	\caption{The measured cross-correlation function $g_{S,aS}^{(N = 12)}$  (red squares) of the 12-mode multiplexed QI as a function of the storage time t. The red curve is the fitting to the measured data according to Eq.(1), where, the data of $R_{inc}^{({\rm{12}})}(t)$  are taken from the Fig.2., together with the parameters ${\rm Z} \approx 2{\rm{ * 1}}{{\rm{0}}^{ - 3}}$,$\bar F \approx 16$  , which come from the measured data, ${\zeta _{se}} = 0.3$  is the fitting parameter.}.
	
	\label{figure4}
\end{figure}

In conclusion, we set up a ring cavity supporting 6 distinguishable TEM$_{00}$ modes and then demonstrated cavity-enhanced and spatially-multiplexed spin-wave-photon quantum interface (QI). The cavity is formed by four flatter mirrors. Two pairs of optical lenses, each of which is set in a confocal structural, are inserted in the cavity. The above arrangement of the cavity makes the mode position in the array to be recovered after a round trip. Also, the cavity arrangement is according to Fermat’s optical theorem\cite{70} , which enable the spatially-multiplexed modes to experience the same optical length per round trip. 

In the presented experiment, the mirror stocked on the piezoelectric transducer is placed at the wait spot of the mode array. Thus, the stabilization of the optical lengths of the multiplexed modes is well achieved by actively driving the piezoelectric transducer. The cold atoms are placed at the confocal position of one pair of the lens, which promises us to achieve effective atom-photon interactions. The spin-wave-photon pairs are generated in the multiplexed modes via DLCZ protocol. BD2 inserted the cavity exclude the ${\sigma ^ - }$- polarized Stokes photon, which is created with magnetic-field-sensitive SWs at the same time, from the cavity modes. Thus, we achieve long-lived storage. The Stokes and anti-Stokes photons in the multiplexed modes are simultaneously resonant with the cavity when the cavity length is locked. The retrieved field from each SW is enhanced via Purcell effect. The average intrinsic retrieval efficiency of the multiplexed SW modes reaches $ \sim $ 70\% at zero delay, together with a storage time of $ \sim $ 600$\mu$s . 

Our presented cavity is different from the reported atom-photon cavity\cite{32, 33, 35, 50,51,52,53}, which use curved mirrors or flatter mirrors + a pair of lenses and support only single modes. Compared with the previous work reported in Ref.\cite{65}, where, the temporally-multiplexed and cavity-enhanced spin-wave–photon QI has been demonstrated, the presented spatially-multiplexed QI doesn’t suffer from additional noise coming from multimode operations. The presented scheme paves a way to make a cavity simultaneously resonant with the mode array, which may be applied in the coupling of spinful bosons to a degenerate multimode optical cavity for simultaneously storing many memories\cite{73}, the coupling of cavity mode array with Rydberg atom array\cite{74} for determinately preparing multiple atom-photon entangled pairs\cite{75}, or in the coupling of a spatial-multimode cavity with nonlinear crystals\cite{37, 76} to prepare multiplexed SPDC photon pairs. Our presented experiment may simultaneously generate multiple spin-wave–photon pairs, which provides a platform to demonstrate connections between repeater nodes. Combining quantum frequency conversion, the Stokes photon frequency can be changed from Rb atomic band into the telecommunications band \cite{11,77,78,79}. The mode number in the presented experiment is 6, which is limited by the size of presented optical elements and can be scaled up by using large-size optical elements in the cavity. Furthermore, if MOT cold atoms are replaced with optical-lattice atoms in our presented scheme, we will obtain a massive multiplexed, high-reversible and long-lived spin-wave–photon QI, which allow us to practically establish quantum repeaters including several long-distance links.

\part*{{\normalsize {\Large Acknowledgements}}}

This work is supported by the National Natural Science Foundation of China (12174235), the Fund for Shanxi Key Subjects Construction (1331), and the Fundamental Research Program of Shanxi Province (202203021221011)

\section{Supplementary material}
\textbf{Cross-correlation function between the Stokes and anti-Stokes fields for the cavity-enhanced spatially-multiplexed atom-photon quantum interface}

 As described in the main text, a write pulse that is applied into the atoms will probabilistically create the correlated pair of a Stokes photon and atomic spin-wave excitation in the 12 modes 1H, 1V, …, iH, … 6V, specially, with photons being in the cavity modes C1H, C1V, …, CiH, … C6V and SWs in the memory modes M1H, M1V, …, MiH, … M6V . In our presented experiment, the detection efficiencies for the various Stokes channels are basically symmetrical, i.e.,${\eta _S}^{({1_H})} \approx {\eta _S}^{({1_V})} \approx ... \approx {\eta _S}^{({i_V})}... \approx {\eta _S}^{({6_V})} \approx {\eta _S}$ . Also, the excitation probabilities $\chi $ for the various Stokes channels are basically symmetrical, which has been explained in the main text. The unconditional probability of detecting a Stokes photon at the detectors DS$i_{\alpha\ }$ (${\alpha\ }$=H,V) is written as: 
 \begin{equation}
 	{P_{D{S_{i\alpha }}}} \approx \chi {\eta _S}
 \end{equation}
In our presented experiment, the detection efficiencies for the various anti-Stokes channels are also basically symmetrical, i.e., ${\eta _{aS}}^{({1_H})} \approx {\eta _{aS}}^{({1_V})} \approx ... \approx {\eta _{aS}}^{({i_V})}... \approx {\eta _{aS}}^{({6_V})} \approx {\eta _{aS}}$  . As explained in our main text, the retrieved anti-Stokes photons in any of spatial modes are routed into the common fiber and detected by the detectors DaS${\alpha\ }$ (${\alpha\ }$=H,V). The unconditional probability of detecting an anti-Stokes photon retrieved from M$i_{\alpha\ }$ SW mode by the detectors may be written as 
 \begin{equation}
	P_{DaS}^{({i_\alpha })} = \chi R_{inc}^{({i_\alpha })}\left( t \right){\eta _{aS}} + \chi \left( {1 - R_{inc}^{({i_\alpha })}\left( t \right)} \right){\xi _{se}}A{\eta _{aS}} + Z{\eta _{aS}}
\end{equation}
where, $R_{inc}^{({i_\alpha })}\left( t \right)$  denotes the intrinsic retrieval efficiency of $i_{\alpha\ }$-th SW mode, Z is the background noise which is assumed to be the same for the different anti-Stokes channels i = 1 to 6 ,$\alpha  = H,V$  ,  ${\xi _{se}}$ denotes the branching ratio corresponding to the read-photon transitions, A is the cavity-enhanced factor (see main text for details). The coincidence probability between Stokes and anti-Stokes fields in $i_{\alpha\ }$-th mode is detected by the detectors DS$i_{\alpha\ }$ and DaS${\alpha\ }$ (${\alpha\ }$=H,V) and written as: 

\begin{equation}
P_{S,aS}^{({i_\alpha })} = \chi R{_{inc}^{({i_\alpha }})}(t){\eta _S}{\eta _{aS}} + {P_{D{S_{i\alpha }}}}P_{DaS}^{({i_\alpha })}
\end{equation}
where, the first term corresponds to quantum correlation counts and the second accidental counts. The across-correlation function of the multiplexed QI storing N=12 modes can be written as:
\begin{equation}
g_{S,aS}^{(N = {\rm{12}})} = \frac{{\sum\limits_{i = 1}^6 {\left( {P_{S,aS}^{({i_H})} + P_{S,aS}^{({i_V})}} \right)} }}{{\sum\limits_{i = 1}^6 {\left( {{P_{D{S_{iH}}}}P_{aS}^{({i_H})} + {P_{D{S_{iH}}}}P_{aS}^{({i_V})}} \right)} }}
\end{equation}
Combining Eqs.2, 3 and 4 into 5, we obtained 
\begin{equation}
g_{S,aS}^{(N = {\rm{12}})} = {\rm{1}} + \frac{{\chi \bar R{{_{inc}^{(12}}^)}(t)}}{{\chi \bar R{{_{inc}^{(12}}^)}(t) + \chi \left( {1 - \bar R{{_{inc}^{(12}}^)}(t)} \right){\xi _{se}}A + Z}}
\end{equation}
where, $\bar R{_{inc}^{(12})}(t)$  is the average intrinsic retrieval efficiency of the 12 SW modes, which equals to the retrieval efficiency $ R{_{inc}^{(12})}(t)$  of the spatially-multiplexed QI storing $N = {\rm{12}}$  SW modes, i.e., $\bar R{_{inc}^{(12})}(t) = R{_{inc}^{(12})}(t)$. So, we have
\begin{equation}
g_{S,aS}^{(N = {\rm{12}})} \approx {\rm{1}} + \frac{{\chi R{{_{inc}^{(12}}^)}(t)}}{{\chi R{{_{inc}^{(12}}^)}(t) + \chi \left( {1 - R{{_{inc}^{(12}}^)}(t)} \right){\xi _{se}}A + Z}}
\end{equation}

\textbf{The measured probabilities of generating spin-wave-photon pairs from the multiplexed QI as function of the number of the stored modes}
We first demonstrate that the multiplexed QI storing  $2 \times m$ SW modes can increase the probability of generating a spin-wave-photon pair per trial by the factor of N, with m being the spatial mode number. The generation probability of a spin-wave-photon pair corresponds to the sum of the Stokes photon detection probabilities at the Stokes detectors DS$i_{H}$ and DSi$i_{V}$ , which is measured as $P_S^{(N)} = \sum\limits_i^m {\left( {{P_{D{S_{iH}}}} + {P_{D{S_{iV}}}}} \right)\;\;} $, where, $P_{D{S_{iH}}_{}}^{} = \chi {\eta _S}$  ( $P_{D{S_{iV}}_{}}^{} = \chi {\eta _S}$ ) denotes the probability of detecting a photon at the detectors  DS$i_{H}$  (DSi$i_{V}$ ) per trial,  ${\eta _S}$ is the detection efficiency for each of Stokes channels, with ${\eta _S}^{({1_H})} \approx {\eta _S}^{({1_V})} \approx ... \approx {\eta _S}^{({i_V})}... \approx {\eta _S}^{({6_V})} \approx {\eta _S}$ assumed. The red squares in Fig.4 are the measured probabilities $P_S^{(N)}$  as a function of the mode number N, which shows that the probability $P_S^{(N)}$    linearly increases with the mode number N.

We also measure the probability of generating the Stokes-anti-Stokes photon pair per trial, which corresponds to coincidence probability $P_{S,aS}^{(N)}$   between the Stokes and anti-Stokes fields. In the measurement, the feed-forward controlled readouts and actively routing of the retrieved photons are utilized, meaning that when a Stokes photon is detected by DS$i_{\alpha\ }$, the read laser pulse is applied on the atoms by a feed-forward controlled AOM and OSN routes the anti-Stokes photon into CSMF after a storage time t. The blue squares in Fig.5 are the measured probabilities  $P_{S,aS}^{(N)}$  as a function of the mode number N, which also shows that the probability  $P_{S,aS}^{(N)}$  linearly increases with the mode number N.
\begin{figure}[h]
	\centering
	\includegraphics[width=3in]{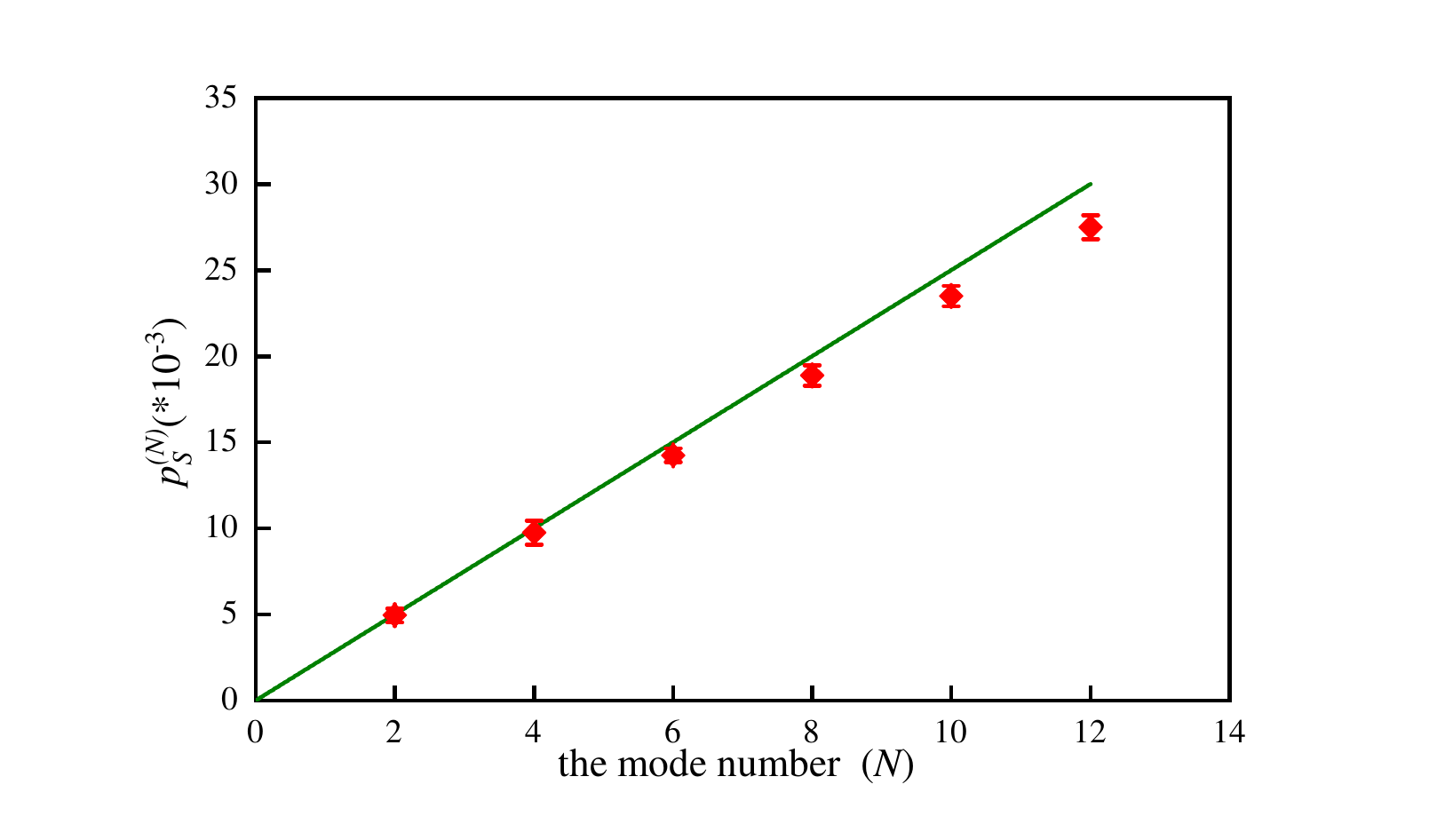}
	\caption{The measured Stokes probability (squares) as a function of the number N of modes. The line is the fitting to the data according to $P_S^{(N)} = N{P_0}$  , with ${P_0} \approx 2.5*{10^{ - 3}}$ corresponding to the single-mode Stokes probability.}.
	
	\label{figure4}
\end{figure}
\begin{figure}[h]
	\centering
	\includegraphics[width=3in]{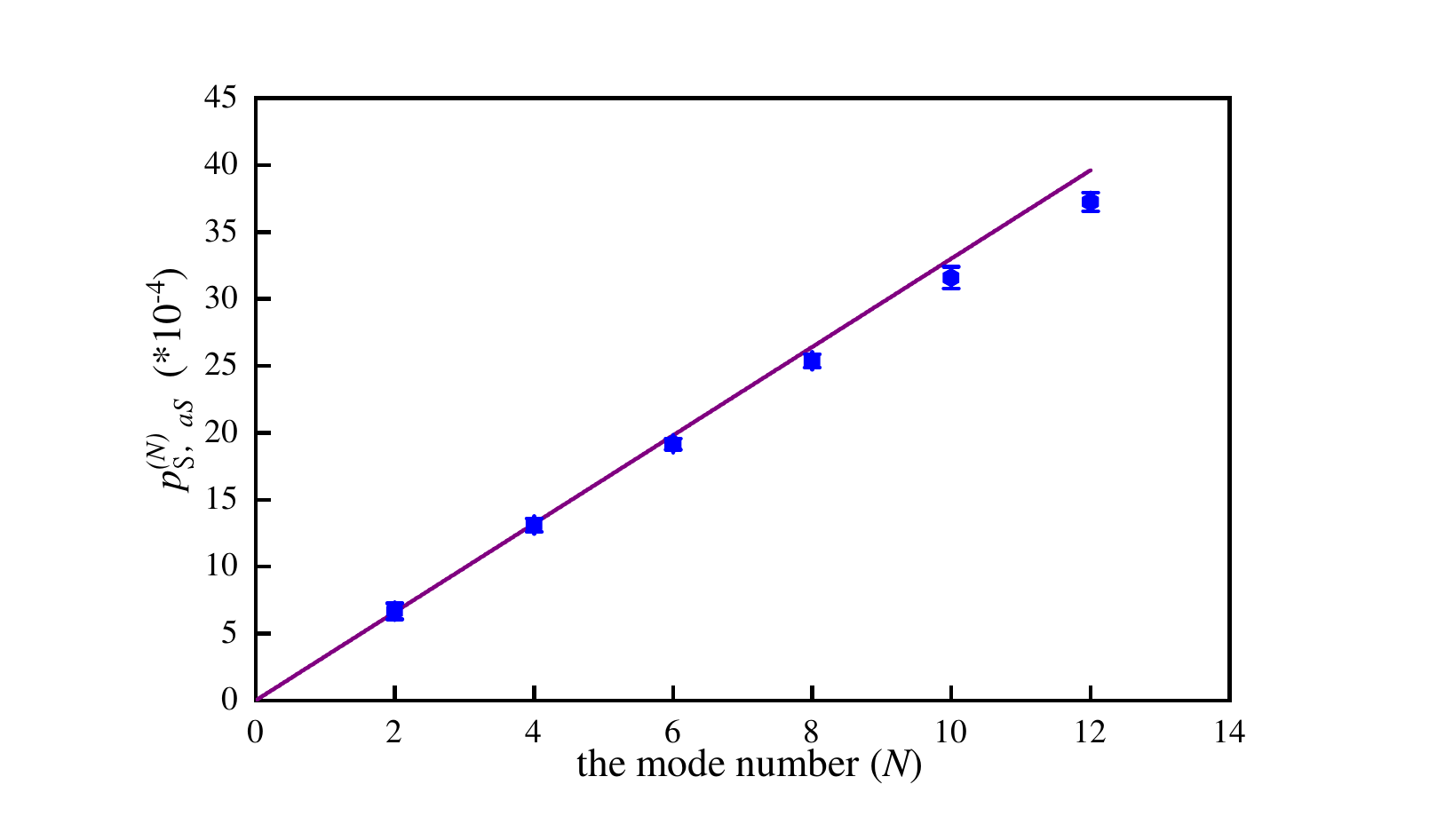}
	\caption{The measured Stokes-anti-Stokes coincident probability (squares) as a function of the mode number 2m. The solid line is the fitting to the data according to $P_{S,aS}^{(N)} = NP_{S,aS}^{(1)}$ , with $P_{S,aS}^{(1)} \approx 3.3*{10^{ - 4}}$ corresponding to Stokes-anti-Stokes coincident probability of an individual mode.
	}.
	
	\label{figure4}
\end{figure}

\end{document}